\documentclass[runningheads]{llncs}
\usepackage{graphicx}
\usepackage{multirow}
\usepackage{calc}
\usepackage{stmaryrd}
\usepackage{tikz}
\usetikzlibrary{shapes.symbols, patterns}
\usepackage{amsmath, amssymb}

\usepackage{mwe}
\usepackage{graphicx}

\usetikzlibrary{shapes}

\usepackage{mathtools}
\usepackage{array}

\usepackage{algorithm,algcompatible,amsmath}
\algnewcommand\INPUT{\item[\textbf{Input:}]}%
\algnewcommand\OUTPUT{\item[\textbf{Output:}]}%
\begin{document}
\title{Student Performance Prediction with Optimum Multilabel Ensemble Model
}
\author{Ephrem Admasu\inst{1*} \and
Abrahaley Teklay\inst{2} }
%\and
%Third Author\inst{3}\orcidID{2222--3333-4444-5555}}
%
\authorrunning{Ephrem A. et al.}
% First names are abbreviated in the running head.1
\institute{
School of Electrical and Computer Engineering, Mekelle University, Ethiopia\inst{1}\\
School of Computing, Mekelle University, Ethiopia\inst{2}
} 
%\and
%Springer Heidelberg, Tiergartenstr. 17, 69121 Heidelberg, Germany
%\url{http://www.springer.com/gp/computer-science/lncs} 
%\and
%\email{ephrem.admasu@mu.edu.et}
%
\maketitle              % typeset the header of the contribution
\begin{abstract}
One of the important measures of quality of education is the performance of students in the academic settings. Nowadays, abundant data is stored in educational institutions about students which can help to discover insight on how students are learning and how to improve their performance ahead of time using data mining techniques. In this paper, we developed a student performance prediction model that predicts the performance of high school students for the next semester for five courses. We modeled our prediction system as a multi-label classification task and used support vector machine (SVM), Random Forest (RF), K-nearest Neighbors (KNN), and Mult-layer perceptron (MLP) as base-classifiers to train our model. We further improved the performance of the prediction model using state-of-the-art partitioning schemes to divide the label space into smaller spaces and use Label Powerset (LP) transformation method to transform each labelset into a multi-class classification task. The proposed model achieved better performance in terms of different evaluation metrics when compared to other multi-label learning tasks such as binary relevance and classifier chains.

\end{abstract}
\section{Introduction}
In recent years, educational data mining (EDM) has been of great research interest due to the abandunce of data about student's information mainly being stored in state databases as well as the increased use of instrumental educational software providng insight on how studnets learn \cite{romero}. The main objective of EDM is to understand and gain knowledge from these educational data using using statistical, machine learning and data mining algorithms and take corrective  measures ahead of time to improve student's perfomance in the educational settings \cite{baker}.

The EDM process follows the same procedure as other application areas in business, medicine, genetics etc. where raw data collected from educational systems is first preprocessed into useful information that could produce insight into the educational system and create awareness of the teaching-learning process \cite{romero2}. Particularly, by analyzing the students' data accurate and efficient student performance prediction models can be designed and developed. Meanwhile, this can help teachers, school administrators, and legal guardians to assist failing students in improving their learning style, organization of resources, effective time management and even address some hindering enviromental or pyschological factors. It also encourages students to take remedial and appropriate actions ahead of time and focus on activities that require high priorities.

In this paper, we developed a multi-label ensemble model to predict high school students' performance based on five courses: English, Math, Physics, Chemistry, and Biology. The dataset for training and testing was collected from three public high schools located in Mekelle, Tigray, Ethiopia. The prediction model evaluates the result of each subject for the next semester as fail or pass, making each label a binary class. The task of prediction is first performed by partitioning the label space L (where $|L| = 5$ for the five courses), into smaller labelsets using community detection methods known as the Stochastic Block Model (SBM) and modularity-maximizing using fast-greedy, and also a randomized partitioning algorithm called Randomized $k$ labELset (RA$k$EL). Then the training data with each labelset is transformed into a single-lable multi-class training set. Each of the single-label classification tasks are trained using a base-level classifier. The base-level classifiers we have considered in this work are Support Vector Machines (SVM), Random Forest (RF), K-Nearest Neighbors (KNN), and Multi-layer Perceptron (MLP).

\section{Literature Review}

In recent years, educational data mining for student performance prediction has gained widespread popularity. Using different techniques and methods, EDM can mine important information regarding the performance of the student and the educational settings to which the students are exposed. Classification, regression, association rules are the most commonly used methods in EDM. Several works have been done for student performance prediction using EDM.

Pandey and Pal \cite{pal} used Bayesian classification method to predict performance of students based on data of 600 students collected from colleges in Awadh University, Faizabad, India. They considered category, language and background qualification of students as input features to predict high and low performing students and take remedial actions for the low performing ones.

With a sample of 300 students, Hijazi and Naqvi \cite{hijazi} predicted student performance using linear regression. In their work, they used attendace, hours spent studying, family income, mother's age and mother's eduction as attributes and showed that mother's education and student's family income are good indicator of student's academic performance.

Shovon M. et al. \cite{shovon} used k-means to predict student performance by grouping them into "Good", "Medium", and "Low" categories. Recent works have also been done using ensemble models. Ashwin Satya Narayana, Marinsz Nuckowski \cite{ashwin} used Decision Trees-J48, Na\"{i}ve Bayes and Random Forest to improve prediction accuracy by removing noisy example in the student's data. They also used combination of rule based techniques such as Apriori, Filtered Associator, and Tertius to identify association rules that affect student outcomes. Natthakan Iam-On et. al. \cite{iam} presented a student dropout prediction model at Mae Fah Fuan University, Thailand using link-based cluster ensemble as a data transformation framework to imporve prediction accuracy. Pooja Kumari et. al. \cite{pooja} used Bagging, Boosting, and Voting Algorithm ensemble methods on Decision Tree (ID3), Nave Bayes, K-Nearest Neighbor, Support vector machines to improve accuracy of student performance prediction. They also showed including student's behavioural (SB) features results in improving the accuracy of the prediction model.

\section{Proposed Work}
\subsection{Objective}
Let $\mathcal{X}$ be an example sapce  which consists of tuples of input values,  discrete and continuous, such that $\forall x_i \in \mathcal{X}, x_i = (x_{i_1}, x_{i_2}, \dots, x_{i_m})$ where $m$ is the number of features and let $\mathcal{L}$ be a label space such that $\mathcal{L} = \{\lambda_1, \lambda_1, \dots, \lambda_L\}$ which is a tuple of $L$ discrete variables of either 0 or 1. $L$ is the number of lables in the dataset. Our training set, $D_{train}$, can be represented as a pair of tuples from the example space $\mathcal{X}$ and label space $\mathcal{L}$ where $D_{train} = \{ (x_i, y_i) \vert x_i \in \mathcal{X}, \, y_i \in \mathcal{L}, 1 \leq i \leq n \}$ and $n$ is the number of examples in the training set i.e. $n = \vert D_{train} \vert$. The goal of multi-label learning model is to find a function $h: \mathcal{X} \rightarrow 2^{\mathcal{L}}$ that maximizes some predictive accuracy or minimizes some loss measure.

After training and validation our model predicts some output $y_i \in  \mathbb{R}^{1 \times L}$ given sample input $x_i \in \mathbb{R}^{1 \times m}$ for some student $student_i$. More precisely, our model predicts a student's results for the next semester in terms of one of the binary classes i.e. Fail or Pass (encoded as 0 and 1, respectively).

\subsection{System Architecture}
The raw data contains sex, age of each student and scores of three consecutive semesters of five courses. It also includes answers to eight closed-ended questions provided by each student. From this, the scores of the third semester are taken to be the output values to be predicted. In the data preprocessing stage, important features are selected from the numerical features (see section 3.3), categorical features (both ordinal and nominal) are mapped into integral values, missing values and outliers are either removed or replaced, the feature input is scaled to give our data the property of a standard normal distribution, and finally the entire data is balanced so that each class is equally represented.

The dataset was then split into two sets: the testing and training sets. The training set was split into separate communities or divisions using partitioning schemes such as RA$k$EL, SBM, and fast-greedy. Each of the partitioned data is then transformed into single-label multi-class classification task using a transformation method known as Label Powerset (LP). Then after, the transformed multi-class data set of each partition was fed into our learning algorithm to train our model. We train our model using different learning algorithms such as SVM, random forest classifier, K-nearest neighbor, and feed forward neural networks.

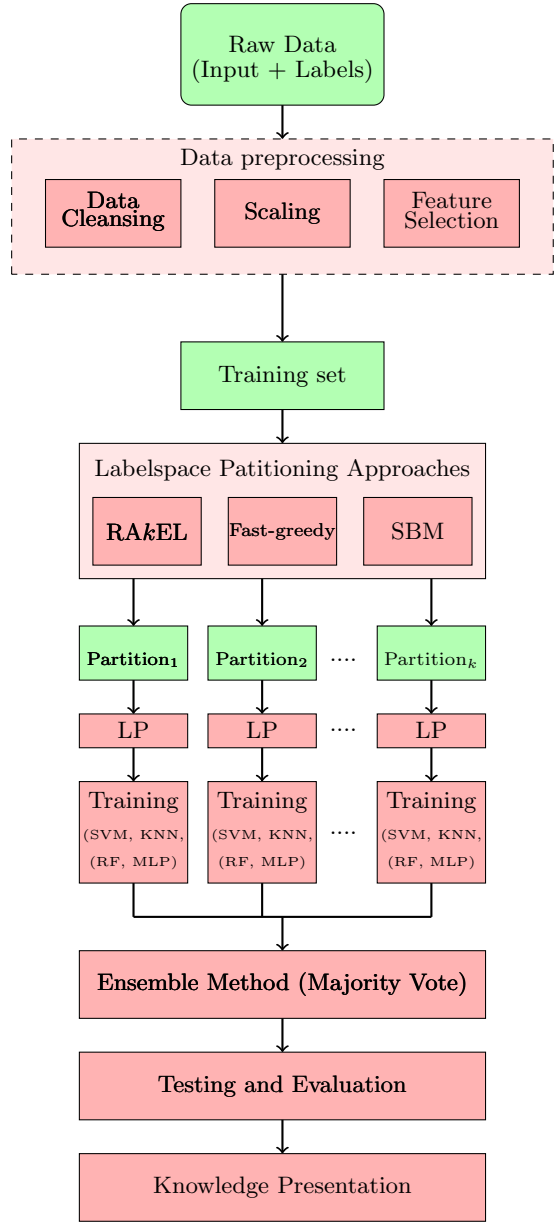
\begin{figure}[!ht]
\begin{center}
\begin{tikzpicture}[scale=.9]

\draw[rounded corners,fill=green!30] (0, 0) rectangle (3, 1.5);
\draw (1.5, 0.9) node{Raw Data};
\draw (1.5, .5) node{(Input + Labels)};

\draw [->, thick] (1.5, 0) -- (1.5, -.5);

\draw[dashed,fill=red!10] (-2.5, -2.5) rectangle (5.5, -.5);
\draw (1.5, -.8) node{\footnotesize Data preprocessing};

\foreach \x in {-1, 1.5, 4} {
\draw[fill=red!30] (\x-1, -1.1) rectangle (\x+1, -2.1);

if(\x==-1){
 \draw (-1, -1.4) node{\footnotesize Data};
\draw (-1, -1.7) node{\footnotesize Cleansing};
}
if(\x==1.5) \draw (1.5, -1.6) node{\footnotesize Scaling};

if(\x==4) {
\draw (4, -1.4) node{\footnotesize Feature};
\draw (4, -1.7) node{\footnotesize Selection};
}

}

\draw[thick, ->] (1.5, -2.5) -- (1.5, -3.5);

\draw[fill=green!30] (0, -4.5) rectangle (3, -3.5);
\draw (1.5, -4) node{Training set};

\draw[thick, ->] (1.5, -4.5) -- (1.5, -5);

\draw[fill=red!10] (-1.5, -7) rectangle (4.5, -5);
\draw (1.5, -5.4) node{Labelspace Patitioning Approaches};

\foreach \x in {-1, 1, 3} {
\draw[fill=red!30] (\x-.3, -6.8) rectangle (\x+1.3, -5.8);

if (\x==-1) \draw (-.5, -6.3) node{RA$k$EL};
else if (\x==1) \draw (1.5, -6.3) node{\scriptsize Fast-greedy};
else \draw (3.5, -6.3) node{SBM};

};

\foreach \x in {-1.5, .4, 2.9} {
\draw[thick, ->] (\x+.8, -7) -- (\x+.8, -7.7);
\draw[fill=green!30] (\x, -8.5) rectangle (\x+1.6, -7.7);

if(\x==-1.5) \draw (-1.5+.8, -8.2) node{\scriptsize Partition$_1$};
if(\x==2.9) \draw (2.9+.8, -8.2) node{\scriptsize Partition$_k$};
if(\x==.4) {
\draw (.4+.8, -8.2) node{\scriptsize Partition$_2$};
}

\draw[thick, ->] (\x+.8, -8.5) -- (\x+.8, -9);
\draw[fill=red!30] (\x, -9.5) rectangle (\x+1.6, -9);
\draw (\x+.8, -9.24) node{LP};

\draw[thick, ->] (\x+.8, -9.5) -- (\x+.8, -10);
\draw[fill=red!30] (\x, -11.5) rectangle (\x+1.6, -10);
\draw (\x+.8, -10.3) node{Training};
\draw (\x+.8, -10.8) node{\tiny (SVM, KNN,};
\draw (\x+.8, -11.2) node{\tiny (RF, MLP)};

\draw[thick] (\x+.8, -11.5) -- (\x+.8, -12);
};
\draw[thick] (-.7, -12) -- (3.7, -12);

\draw (2.4, -8.2) node{....};
\draw (2.4, -9.24) node{....};
\draw (2.4, -10.75) node{....};

\foreach \y in {-12.5, -14, -15.5}{

\draw[thick,->] (1.5, \y+.5) -- (1.5, \y);
\draw[fill=red!30] (-1.5, \y-1) rectangle (4.5, \y);

if(\y==-12.5){
\draw (1.5, -13) node{Ensemble Method (Majority Vote)};
}
if(\y==-14)
\draw (1.5, -14.5) node{Testing and Evaluation};
if(\y==-15.5)
\draw (1.5, -16) node{Knowledge Presentation};
};
\end{tikzpicture}
\end{center}
\caption{\label{fig_1} Scheme of the proposed system}
\end{figure}

After training, the trained model was tested using the testing set which also went through the same partitioning and transformation procedures. The output of each label was predicted using majority voting rule since one label can exist in more than partition. The majority voting rule is an ensemble method which takes the majority value to predict the output of a given label. Finally, the performance of the algorithm was evaluation using different evaluation measures. The overall architecture of the system is shown in Figure \ref{fig_1}.

\subsection{Dataset}
The dataset was gathered from three public high schools located in the city of Mekelle, Ethiopia. The process of data collection was divided into two separate tasks. First, basic information such as student name, ID, sex and their scores on five courses i.e. English, Mathematics, Physics, Biology and Chemistry, of three consecutive semesters was collected from the school adminstrators. Second, a questionaire was distributed to all students. This includes questions such as students perception towards the importance of education, family's educational background, family average income, student's grade 10 GPA score and others. Table \ref{tab_data} summarizes all the variables of dataset.

\begin{table}
\caption{Dataset features with possible values}\label{tab_data}
\begin{tabular}{
|p{.15\textwidth-3\tabcolsep}|
p{.2\textwidth-3\tabcolsep}|
p{.35\textwidth-3\tabcolsep}|
p{.3\textwidth-3\tabcolsep}|
}
\hline
Feature &  Type & Description & Possible values \\
\hline
Gender & Categorical (nominal) & Gender of the student & Male or Female \\
\hline
Age & Numerical & A positive integer value & Range: $[15, 21]$ \\
\hline
\multirow{2}{*}{Courses} &  \multirow{2}{*}{Numerical} & Courses include English  & \\
 & & Math, Physics, Chemistry, Biology & $[0, 100]$ \\
% & & Chemistry and Biology & Very Good: $[70, 85)$ \\
% & & & Good: $[60, 70)$ \\
% & & & Satisfactory: $[50, 60)$ \\
% & & & Fail: $[0, 50)$ \\
\hline
Quality of education & Categorical (nominal) & Student's perception on the quality of education in his/her high school & \{Excellent, Very good, Good, Satisfactory, Bad, Very bad\} \\
\hline
Legal guardians & Categorical (nominal) & Parents/guardians who take care of the student & \{mother and father, father only, mother only, siblings, other, live alone\} \\
\hline
Family income & Categorical (ordinal) & Total average income of family/guardians & \{$<$5000, 5000-10000, 10000-20000, $>$20000\}\\
\hline
Family educational background & Categorical (ordinal) & Family's highest education background & \{diploma, degree, masters, PhD, high school, high school dropout, no education\}\\
\hline
Tutorial & Categorical (nominal) & Does the student get extra help, such as tutorial? & \{Yes, No\} \\
\hline
Grade 10 GPA & Categorical (ordinal) & Student's GPA in grade 10 & \{2-2.5, 2.5-3, 3-3.5, 3.5-4\}\\
\hline
Parent occupation & Categorical (nominal) & Student guardian's occupation & \{Civil servant, Artisan, Trading/merchant, Military\}\\
\hline
Student's perception towards eduction & Categorical (nominal) & Does the student think education is helpful for his future? & \{Yes, No\} \\
\hline
\end{tabular}
\end{table}

As is depicted in Table \ref{tab_data} the features of the dataset can be numerical or categorical. Numerical features represent real numbers where as categorical features are further divided into two: nominal and ordinal. Ordinal features are categorical values that can be ordered or sorted where as in nominal features order is not inherent. Hence, family income, family education background, grade 10 scores are ordinal features and gender, student's perception on the quality of education, legal guardians, tutorial, family occupation, and student's perception on the importance of education are nominal features. 

After preparing the dataset, the next step is to separate the input features and the target (output) values. In this study, we are building a machine learning model that predicts the performance by predicting a student's score in the next semester. We have collected scores of five courses within three consecutive semesters. Then, we used the scores of the first two semsters as input features, along with the other features, to predict the results of the third semester. Since we are building a classification model, we discretized the scores into two classes, 0 and 1, using two interval values $>=$50 or $<$50, respectively. After preprocessing, the total size of the dataset was 893 and 70\%, i.e 625, was used for training and the rest 268 for testing.

\subsection{Feature Selection and Transformation}
Feature selection is an important part of builindg an good machine learning model. One of the most common ways to select features is to find the correlation between the numerical input features and the output values using Peason's correlation coefficient $r$ given as:

\begin{equation}
r = \frac{\sum_{i=1}^n (X_i - \bar{X})(Y_i - \bar{Y})}{\sqrt{\sum_{i=1}^n (X_i - \bar{X})^2}\sqrt{\sum_{i=1}^n (Y_i - \bar{Y})^2}}
\end{equation}
where $X$ and $Y$ are random variables and $\bar{X}$ and $\bar{Y}$ mean of $X$ and $Y$ respectively. The of value of $r$ is always between -1 and 1. The higher the magnitude of $r$, is a good indicator that one of the two variables can be used as a feature input to the other variable.

Our dataset contains 11 numerical features; 10 features are scores of the five courses from previous two semesters and one feature is the age of a student. By finding the correlation of each of these features with five output values, we can understand the relationships between the feature input and the target values. For instance, table \ref{tab_corr} shows the correlation of the input feature with the scores of Math course.

\begin{table}
\begin{center}
\caption{Correlation between input and output varialbes (Note: for a given course $C$, $C_i$ represents course scores at semester $i$)}\label{tab_corr}
\begin{tabular}{
|p{.15\textwidth-3\tabcolsep}|
p{.15\textwidth-3\tabcolsep}|
p{.15\textwidth-3\tabcolsep}|
p{.15\textwidth-3\tabcolsep}|
p{.15\textwidth-3\tabcolsep}|
p{.15\textwidth-3\tabcolsep}|
}
\hline
Feature &  Eng\_3 & Math\_3 & Phy\_3 & Chem\_3 & Bio\_3 \\
\hline
Eng\_1 & 0.492808 & 0.404363 & 0.440780 & 0.329905 & 0.259480 \\
\hline
Eng\_2 & 0.552929 & 0.449800 & 0.494027 & 0.387672 & 0.438640 \\
\hline
Math\_1 & 0.416950 & 0.468981 & 0.545903 & 0.484068 & 0.276077 \\
\hline
Math\_2 & 0.331981 & 0.431253 & 0.514556 & 0.403260 & 0.254936 \\
\hline
Phy\_1 & 0.397704 & 0.329259 & 0.493856 & 0.442832 & 0.266141 \\
\hline
Phy\_2 & 0.443926 & 0.381799 & 0.498592 & 0.411874 & 0.359873 \\
\hline
Chem\_1 & 0.443961 & 0.545993 & 0.531688 & 0.501924 & 0.341021 \\
\hline
Chem\_2 & 0.416666 & 0.528435 & 0.517878 & 0.426985 & 0.316601 \\
\hline
Bio\_1 & 0.408049 & 0.369690 & 0.510147 & 0.404698 & 0.258837 \\
\hline
Bio\_2 & 0.231319 & 0.187608 & 0.398980 & 0.402353 & 0.238710 \\
\hline
\end{tabular}
\end{center}
\end{table}

From table \ref{tab_corr} we can observe the following points:
\begin{enumerate}
\item Intuitively, one would think that for a given course, the scores from two previous semesters of the same course would have a higher correlation than the other course scores. However, table \ref{tab_corr} tells us that this is almost always not the case. For example, there is a higher correlation between Math\_3 and Chem\_1 and Chem\_2 than Math\_3 and Math\_1 and Math\_2.
\item Almost all the correlation values in the table are $> 0.3$. This indicates that all input values are good candidates to be used input features.
\end{enumerate}

Since ordinal features have order among their values, we mapped each ordinal feature $f_i$ with $n$ unique values into a set of integers $i \in \{1, \dots, n\}$ where each feature value is assigned a number based on its order in the feature. For nominal features we used one-hot encoding which creates a new dummy feature for each unique value in the nominal feature column \cite{onehot}.

\section{Multi-label Ensemble Model}
Our dataset contains five output labels which makes it a multi-label classification task. When input instances are allocated only one category, the task becomes single-label classification task. The field of single-label classification is more matured than multi-label classification for solving classification problems. Therefore, in most cases the multi-label classification is transformed into a single-label classification problem through the so-called \textit{problem transformation} methods. The two commonly used transformation methods are Label powerset (LP) and Binary relevance. In this work, we use label powerset.

\textit{Label powerset (LP)} is a transformation method that maps each unique label combination into a unique class. One strength of LP is that it preserves the correlation between labels. The weakness of LP is that mapping a label space of size $\vert L \vert$ requires yielding a multi-class classification problem of $2^{\vert L \vert}$ classes which can be impractically large as $\vert L \vert$ gets bigger. One way to circumvent this problem is to limit the unique label combinations only to the ones that occur in training set but this also has a problem of overfitting and a only small number of training examples are associated with most of the classes \cite{lp,rakel}.

We can avoid the problems of LP by partitioning the label space into smaller labelsets and apply LP in the labelsets. We consider three common ways to partition the labelspace: 1) RAndom $k$ labELsets (RA$k$EL) 2) Data-driven partioning and 3) Stochastic Block Model (SBM). RA$k$EL divides the number of labels into smaller labels by randomly picking the label groups. Here, $k$ denotes the number of the labelsets. RA$k$EL comes in two variants: RA$k$EL$_d$, which partitions the labels into $k$ \textit{disjoint} labelsets, and RA$k$EL$_o$, which also partitions the labels into $k$ labelsets but allows \textit{overlapping} of label subspaces.

The second method of label partitioning is based on acquiring relationships in the training set and constructs a graph, so-called \textit{Label co-occurrence graph}, using the training data and creates partitions in such a way that the modularity is maximized. Modularity is a measure of strength of partitioning a graph into modules or clusters. A detailed treatment of these concepts is found in \cite{datadriven,newman}. Algorithms for partitioning the label co-occurence graph by maximizing the modularity measure include fast-greedy \cite{greedy}, leading eigenvector approximation method \cite{eigenvector}, infomap \cite{infomap}, label propagation algorithm \cite{propagation}, and walktrap \cite{walktrap}.

\subsection{Label Partitioning with Stochastic Block Model}

Stochastic Block Model (SBM) infers or recovers $b$ partitions (blocks) from the graph constructed using the training set. SBM is formulated as follows. Given the number of partitions $b$, the number of vertices in a graph $n$, a probability vector $p$ of dimension $b$, and a symmetric matrix $W$ of dimension $k \times k$ with entries in $[0, 1]$, the model partitions the vertices of the graph into $b$ communities, where each vertex is assigned a division label in $\{1, \dots, b\}$ independently under the community prior $p$, and pairs of vertices with labels $i$ and $j$ connect independently with probability $W_{i,j}$. Many algorithms have been proposed for community detection in SBM \cite{abbe,sbm}.

\section{Results and Discussion}
\subsection{Environmet}
In this paper, we used \texttt{scikit-multilearn} \cite{skmul} -- a scikit-learn API compatible library for multi-label classification in python which supports several classifiers and label partition models. We have also used \texttt{scikit-learn} \cite{scikit} for data preprocessing and evaluation metrics. \texttt{scikit-learn }is widely used in the scientific Python community and supports many machine learning application areas.

\subsection{Evaluation Metrics}
The evaluation metrics used for single-label classification are different from multilabel classification. In single-label classification the training samples can be either correct or incorrect. In multil-lable classification since labels introduce additional degrees of freedom it is important to consider multiple and contrasting measures \cite{madjarov}. In this study, we use three example-based measures: accuracy, Hamming loss, and Jaccard similarity as well as one label-based measure, F1, evaluated by two averaging schemes: micro and macro. We also use the following definitions as discussed in \cite{datadriven}:
\begin{itemize}
\item $X$ is the set of objects used in the testing scenario for evaluation
\item $L$ is the set of labels that spans the output space $Y$
\item $\bar{x}$ denotes an example object undergoing classification
\item $h(\bar{x})$ denotes the label set assigned to object $h(\bar{x})$ by the evaluated classifier $h$
\item $y$ denotes the set of true labels for the observation $\bar{x}$
\item $tp_j$, $fp_j$, $fn_j$, $tn_j$ are respectively true positives, false positives, false negatives and true negatives of the of label $L_j$, counted per label over the output of classifier h on the set of testing objects
$\bar{x} \in X$, i.e., $h(X)$
\item the operator $\llbracket p \rrbracket$ converts logical value to a number, i.e. it yields 1 if p is true and 0 if p is false
\end{itemize}

The example-based metrics, Hamming loss, subset accuracy, and Jaccard similarity and the label-based metric, F1 measure, are defined as follows:
\subsubsection{\textbf{1) Hamming Loss}:} evaluates the number of times an example-label pair is misclassified, i.e., label not belonging to the example is predicted or a label belonging to the example is not predicted. $\otimes$ denotes the logical exclusive or.
\begin{equation}
HammingLoss(h) = \frac{1}{\vert X \vert}\sum_{\bar{x} \in X} \frac{1}{\vert L \vert}\llbracket (L_j \in h(\bar{x})) \otimes (L_j \in y) \rrbracket
\end{equation}
\subsubsection{\textbf{2) Accuracy score (Subset accuracy)}}:  is intance-wise measure that evaluates the set of predicted labels for a sample that exactly match the corresponding set of true labels. 
\begin{equation}
SubsetAccuracy(h) =  \frac{1}{\vert X \vert}\sum_{\bar{x} \in X} \llbracket h(\bar{x}) = y \rrbracket
\end{equation}

\subsubsection{\textbf{3) Jaccard Similarity}:} also simply called accuracy, is a measure of similarity between the predicted and true lables. It evaluates the coefficient of the size of the intersection between the predicted and true labels and size of their union.

\begin{equation}
Jaccard(h) =  \frac{1}{\vert X \vert}\sum_{\bar{x} \in X} \frac{h(\bar{x}) \cap y}{h(\bar{x}) \cup y}
\end{equation}

\subsubsection{\textbf{4) F1 Measure}:} The label-based evaluation method we use in this work is the F1 measure. F1 measure is the harmonic mean of precision and recall and is often considered to be a good indicator of the relationship between precision and recall. Precision is the measure of how much negative cases are misclassified as positives and recall is the measure of how much positive cases are misclassified as negatives. The average of these two measures is computed using two different methods, micro- and macro-averaging, which can give different interpretations specially in multi-labels settings. 

Micro-averaging takes the aggregate contributions of all classes true/false positives/negatives and computes the average metric. This is given as:

\begin{equation}
\begin{split}
\mbox{precision}_{micro}(h) = \frac{\sum_{j=1}^{\vert L \vert} tp_j}{\sum_{j=1}^{\vert L \vert} tp_j + fp_j}  \\
\mbox{recall}_{micro}(h) = \frac{\sum_{j=1}^{\vert L \vert} tp_j}{\sum_{j=1}^{\vert L \vert} tp_j + fn_j} \\
\mbox{F1}_{micro}(h) = 2\cdot\frac{\mbox{precision}_{micro}(h) \cdot \mbox{recall}_{micro}(h)}{\mbox{precision}_{micro}(h) + \mbox{recall}_{micro}(h)}
\end{split}
\end{equation}

On the other hand, macro-averaging first evaluates the metric independently for each class and takes the average over the number of labels. Hence, macro-averaging treats all classes equally while micro-averaging does not which makes suitable for cases where there is class imbalance.

\begin{equation}
\begin{split}
\mbox{precision}_{macro}(h, j) = \frac{tp_j}{ tp_j + fp_j}  \\
\mbox{recall}_{macro}(h, j) = \frac{tp_j}{tp_j + fn_j} \\
\mbox{F1}_{macro}(h, j) = 2\cdot\frac{\mbox{precision}_{macro}(h, j) \cdot \mbox{recall}_{macro}(h, j)}{\mbox{precision}_{macro}(h, j) + \mbox{recall}_{macro}(h, j)}
\end{split}
\end{equation}

\subsection{Evaluation results}
In this study, we used four different base-level classifiers and compared the results of the student performance using the evaluation metrics discussed in section 5.2. The four base-level classifiers are SVM, Random forest (RF), KNN, and Multilayer perceptron (MLP). We also use three partitioning approaches to partition the label space of the multi-label classifier and employ LP to train the corresponding classifier with partitioned labelsets. The three label partition methods are RA$k$EL$_o$ (RAndom $k$ labELsets with overlapping), data-driven parition using fast greedy, and stochastic block model (SBM). Note that there are other variants of RA$k$EL, known as RA$k$EL$_d$, and modularity maximizing algorithms (discussed in section 4), but we consider  RA$k$EL$_o$ and fast-greedy because RA$k$EL$_o$ generated better results in our dataset and fast-greedy outputed the same result as its variants. For each classifier we ran several tests and chose the output with the best performance.

\subsubsection{1) Evaluation results with different label partition approaches:} 
Table \ref{fast_greedy}, \ref{rakelo}, and \ref{sbm} show the evaluation results of the student performance  multi-label classifier model using the four base-level classifiers. The results of classifier with the best performance are bolded and to the average best results across different tables are underlined.

Table \ref{fast_greedy} uses fast-greedy to partion the student dataset label space. As is shown in the table, SVM achieved the best performance in terms of all the evaluation metrics getting 19.5\% Hamming loss, 69.2\% Jaccard similarity, 33.8\% subset accuracy, 82.2\% F1 micro, and 81.3\% F1 macro. 

\begin{table}[h]
\caption{Evaluation results for four base classifiers with fasty-greedy approach to partition the label space}\label{fast_greedy}
\begin{center}
\begin{tabular}{
p{.16\textwidth-3\tabcolsep}
p{.13\textwidth-3\tabcolsep}
p{.13\textwidth-3\tabcolsep}
p{.13\textwidth-3\tabcolsep}
p{.13\textwidth-3\tabcolsep}
p{.13\textwidth-3\tabcolsep}
}

\textbf{Metric}  & \textbf{SVM} & \textbf{RF} & \textbf{KNN} & \textbf{MLP} & \textbf{Avg.}\\\hline
Hamming. & \textbf{0.195} & 0.218 & 0.243 &  0.223 & 0.220 \\
Jaccard. & \textbf{0.692} &  0.674 &  0.679 & 0.678 & 0.681 \\
Subset Acc. & \textbf{0.338} & 0.327 & 0.241 & 0.334 & 0.310 \\
F1$_{micro}$ & \textbf{0.822} & 0.796 & 0.794 & 0.804 & 0.804 \\
F1$_{macro}$ & \textbf{0.813} &  0.784 & 0.779 & 0.791 & 0.792 \\
\end{tabular}
\end{center}
\end{table}
Table \ref{rakelo} shows the evaluation results of the multilabel classifier with the four base-label classfiers with the lable space of the student performance dataset partioned using RA$k$EL$_o$. Note that the number of labelsets $k$ for RA$k$EL$_o$ is chosen to be 5, and the cluster size is 3 because these are the optimum values we found using brute force. The table shows SVM to outperform the rest of base-classifiers in terms of Hamming loss (22.4\%), micro F1 (83.3\%) and macro F1 (82.4\%) measures whereas Random forest present the best performance in terms of subset accuracy with 34.4\% and MLP demonstrated the best perormance in terms of Jaccard similairty with 72.3\%. Similarly, in comparison to the base-classifiers in table \ref{fast_greedy} and table \ref{rakelo}, we can observe that SVM, Random forest, and MLP achieved better results in terms of all evaluation metrics except Hamming loss when RA$k$EL$_o$ partioning is used instead of fast-greedy approach. Furthermore, KNN achieved better results in terms of all evaluation metrics except Jaccard similarity when RA$k$EL$_o$ partitioning is used instead of fast-greedy approach. In conclusion, when the student performance multilable classification task is partitioned with RA$k$EL$_o$ instead of fast-greedy method, it achieved better results in most evaluation metrics with all the base-level classifiers.

\begin{table}
\caption{Evaluation results for four base classifiers with RA$k$EL$_o$ approach to partition the label space}\label{rakelo}
\begin{center}
\begin{tabular}{
p{.16\textwidth-3\tabcolsep}
p{.13\textwidth-3\tabcolsep}
p{.13\textwidth-3\tabcolsep}
p{.13\textwidth-3\tabcolsep}
p{.13\textwidth-3\tabcolsep}
p{.13\textwidth-3\tabcolsep}
}
\textbf{Metric}  & \textbf{SVM} & \textbf{RF} & \textbf{KNN} & \textbf{MLP} & \textbf{Avg.}\\\hline
Hamming. & \textbf{0.224} & 0.238 & 0.227 & 0.226 & 0.229 \\
Jaccard. & 0.711 &  0.704 &  0.649 & \textbf{0.723} & 0.697 \\
Subset Acc. & 0.342 & \textbf{0.344} & 0.316 & 0.337 & 0.335 \\
F1$_{micro}$ & \textbf{0.833} & 0.829 & 0.831 & 0.829 & \underline{0.831} \\
F1$_{macro}$ & \textbf{0.824} &  0.819 & 0.819 & 0.816 & 0.820 \\
\end{tabular}
\end{center}
\end{table}

Table \ref{sbm} presents the summary of evaluation results for student performance multilable classifier when the lable space is partitioned with SBM. As can be seen in the table, SVM outperformed the other classifiers in terms of Jaccard similarity(73.7\%) and micro F1 (84.2\%)  and Random forest recorded the best performance in terms of Hamming loss (19.1\%) and subset accuracy (40.2\%) compared to the other base classifiers. In fact, Random forest with SBM partitioning has the lowest Hamming loss and the highest subset accuracy comapred to any base-level classifier with any partition scheme and the same is true for SVM considering Jaccard similarity and micro F1. MLP also achieved the highest macro F1 (87.2\%) when used as a base-level classfier for a multi-class data partitioned by SBM. The average evaluation results in table \ref{sbm} are also the best, except for micro F1,  results compared to that of tables \ref{fast_greedy} and \ref{rakelo}.

\begin{figure}
\begin{center}
\includegraphics[scale=.5]{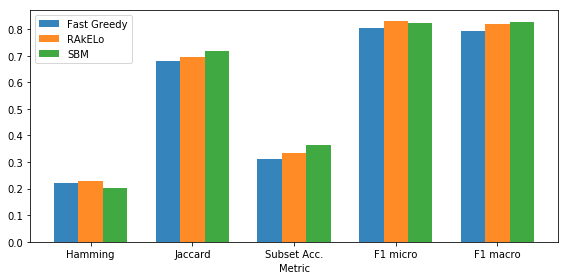}
\end{center}
\caption{\label{comp_fig} Comparison of averaged evaluation results for the three partition methods}
\end{figure}

\begin{table}
\caption{Evaluation results for  four base classifiers with SBM approach to partition the label space}\label{sbm}
\begin{center}
\begin{tabular}{
p{.16\textwidth-3\tabcolsep}
p{.13\textwidth-3\tabcolsep}
p{.13\textwidth-3\tabcolsep}
p{.13\textwidth-3\tabcolsep}
p{.13\textwidth-3\tabcolsep}
p{.13\textwidth-3\tabcolsep}
}
\textbf{Metric}  & \textbf{SVM} & \textbf{RF} & \textbf{KNN} & \textbf{MLP} & \textbf{Avg.}\\\hline
Hamming. & 0.201 & \textbf{0.191} & 0.225 & 0.201 & \underline{0.204} \\
Jaccard. & \textbf{0.737} &  0.730 &  0.712 & 0.693 & \underline{0.718} \\
Subset Acc. & 0.378 & \textbf{0.402} & 0.322 & 0.361 & \underline{0.366} \\
F1$_{micro}$ & \textbf{0.842} & 0.823 & 0.817 & 0.821 & 0.825 \\
F1$_{macro}$ & 0.823 &  0.812 & 0.802 &\textbf{0.872} & \underline{0.827} \\
\end{tabular}
\end{center}
\end{table}
Figure \ref{comp_fig} presents the averaged evaluation results across all the base-level classifiers with the four evaluation metrics. From the figure we can observe that multi-label classification with SBM partition produced better results in terms of Hamming accuracy, Jaccard similarity, subset accuracy, and macro F1. It also shows that SBM-partitioned multilable student performance label space  would generate almost the same results in terms of micro F1 comapred to RA$k$AL$_o$.

\subsubsection{2) Evaluation results with different problem transformation methods:} So far, we used LP multi-learning method which transforms each partitioned label space that exists in the student performance training set into a single-label multi-class classification task. Moreover, we can consider other problem transformation methods such as Binary relevance \cite{overview} and classifier chain \cite{cc}. Table \ref{br} and \ref{cc} show the evaluation results of the four base-classifiers using binary relevance and classifier chains as the multi-label learning methods. Note that we used SBM as our label space partitioning scheme due better performance compared to the other schemes discussed in the previous subsection.

\begin{table}
\caption{Evaluation results for four base classifiers with Binary relevance multi-learning method}\label{br}
\begin{center}
\begin{tabular}{
p{.16\textwidth-3\tabcolsep}
p{.13\textwidth-3\tabcolsep}
p{.13\textwidth-3\tabcolsep}
p{.13\textwidth-3\tabcolsep}
p{.13\textwidth-3\tabcolsep}
p{.13\textwidth-3\tabcolsep}
}
\textbf{Metric}  & \textbf{SVM} & \textbf{RF} & \textbf{KNN} & \textbf{MLP} & \textbf{Avg.}\\\hline
Hamming. & \textbf{0.217} & 0.247 & 0.258 & 0.235 & 0.239 \\
Jaccard. & \textbf{0.708} &  0.694 &  0.685 & 0.675 & 0.685 \\
Subset Acc. & \textbf{0.337} & 0.310 & 0.268 & 0.289 & 0.301 \\
F1$_{micro}$ & \textbf{0.824} & 0.790 & 0.791 & 0.803 & 0.802 \\
F1$_{macro}$ & \textbf{0.816} &  0.773 & 0.781 & 0.794 & 0.793 \\
\end{tabular}
\end{center}
\end{table}

Table \ref{br} shows the performance of the four base-level classifiers when binary relevance is used as a multi-learning method. The table depicts that SVM outperformed all the other classifiers in all evaluation metrics. When compared to table \ref{sbm}, however, the results show that using LP has better performance in terms of all the measures when compared to using binary relevance.

\begin{table}
\caption{Evaluation results for four base classifiers with Classifier chains as multi-learning method}\label{cc}
\begin{center}
\begin{tabular}{
p{.16\textwidth-3\tabcolsep}
p{.13\textwidth-3\tabcolsep}
p{.13\textwidth-3\tabcolsep}
p{.13\textwidth-3\tabcolsep}
p{.13\textwidth-3\tabcolsep}
p{.13\textwidth-3\tabcolsep}
}
\textbf{Metric}  & \textbf{SVM} & \textbf{RF} & \textbf{KNN} & \textbf{MLP} & \textbf{Avg.}\\\hline
Hamming. & \textbf{0.217} & 0.226 & 0.259 & 0.224 & 0.232 \\
Jaccard. & \textbf{0.711} &  0.692 &  0.665 & 0.698 & 0.692 \\
Subset Acc. & \textbf{0.345} & 0.324 & 0.282 & 0.316 & 0.317 \\
F1$_{micro}$ & \textbf{0.832} & 0.819 & 0.797 & 0.822 & 0.818 \\
F1$_{macro}$ & \textbf{0.818} &  0.801 & 0.781 & 0.808 & 0.802 \\
\end{tabular}
\end{center}
\end{table}

Table \ref{cc} presents the performance evaluation of the student performance testing set with base-level classifiers and classifier chains as a multi-label method. From the table we can see that SVM outperforms all the other base-classifiers in terms all the evalautions measures, a similar result as in table \ref{br}. Comparing tables \ref{cc} and \ref{sbm}, we can see that the performance of multi-label classification with LP as a multi-label learning is better than classifier chains in all base-classifiers in terms of almost all evaluation measures. 

\begin{figure}[h]
\begin{center}
\includegraphics[scale=.5]{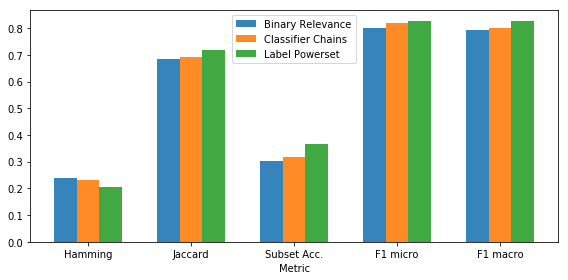}
\end{center}
\caption{\label{comp2} Comparison of averaged evaluation results for the three problem transformation methods with label space partitioned with SBM}
\end{figure}

Figure \ref{comp2} shows the comparison of the averaged evaluation results of the five base-classifiers using binary relevance, classifier chains, and label powerset for the student performane data set. Label powerset outperforms all the other methods in all evaluation methods. Therefore, we can conclude that the student performance multilable learning task generates the best results when SBM is used as partitioning approach and LP is used as the problem transformation method.

\section{Conclusion and recommendation}

This paper has presented a student performance prediction using a multilabel learning method that learns an ensemble of LP classifiers where each classifiers train a subset of the set of labels that are partioned using SBM. The partitioning method was compared to the stat-of-art approaches called RA$k$EL$_o$, which allows overlapping of the subset labels, and modularity-maxing community-detection using fast-greedy, which is a data-driven partition scheme. The evaluation results conducted on four base-classifiers show that the student prediction performance model generated better results when SBM is used to partition the label space of the student's dataset. Furthermore, the LP ensemble model was compared to other well-known problem transformation methods, binary relevance and class chains. The ensemble of LP classifiers produced better results in terms of different evaluation schemes than binary relevance and class chains.

As a future work, we will evaluate the proposed multi-lable ensemble model on student dataset with more training samples and higher label spaces. Specifically, the model can produce pronounced results if the dataset and label space are of larg sizes. Therefore, we will consider using this model to predict student's performance particularly for matriculation exams of 8$^{th}$, 10$^{th}$, and 12$^{th}$ grades.
\bibliographystyle{splncs04}

\end{document}